\documentclass[twocolumn,showpacs,preprintnumbers,amsmath,amssymb,pra]{revtex4-1}
\usepackage{graphicx}
\usepackage{dcolumn}
\usepackage{ulem}
\usepackage{textcomp}
\usepackage{bm}
\usepackage{color}

\newcommand{\ord}{{\cal O}}

\newcommand{\bea}{\begin{eqnarray}}
\newcommand{\ea}{\end{eqnarray}}
\begin{document}


\title{Robustness of fragmented condensate  many-body states for continuous distribution amplitudes in Fock space}

\author{Uwe R. Fischer and Bo Xiong}
\affiliation{Seoul National University,  Department of Physics and Astronomy \\ Center for Theoretical Physics, 151-747 Seoul, Korea}
\date{\today}

\begin{abstract}
We consider a two-mode model describing scalar bosons with two-body interactions in a single trap, taking into account coherent
pair-exchange between the modes.
It is demonstrated that the resulting fragmented many-body states with continuous (nonsingular) Fock-space distribution amplitudes are robust against perturbations due to occupation number and relative phase fluctuations, 
Josephson-type tunneling between the modes, and weakly broken parity of orbitals,  
as well as against perturbations due to interaction with a third mode.
\end{abstract}

\pacs{03.75.Hh, 
03.75.Nt, 	
03.75.Lm 	
}

\maketitle
     
   \section{Introduction}

Conventional wisdom has it that fragmented condensates, i.e., many-body states leading to more than one macroscopic 
eigenvalue of the single-particle density matrix \cite{penr}, are unstable against small perturbations when contained in a single (e.g., harmonic) trap \cite{muel,jack}. 
It is well established that stable fragmentation can be readily prepared for spatially well-separated 
modes in the field operator expansion, for example in deep double wells \cite{spek} or in an optical lattices \cite{jaks,kase,grei}.  On the other hand, fragmented condensate states in a single trap 
are conventionally obtained around special points of symmetry of the system, e.g., in spin-1 Bose gases \cite{ho, koas}, rotating gases \cite{wilkin,rokhsar,ueda,liu}, and spin-orbit coupled systems \cite{zhou}. These fragmented condensate many-body states obtained from symmetry in a single trap are sharply peaked in Fock space; as a consequence, they are inherently unstable against
perturbations and decay into single condensates. 
Recent theoretical work however put forward the possibility of \textit{robust} fragmented condensate states in a single trap, 
with significant (that is, not exponentially small) spatial overlapping of the field operator modes \cite{phil}. 
The corresponding class of fragmented condensate many-body states was subsequently shown to be immune against perturbations 
on the dynamical many-body level, i.e., under rapid changes of interaction couplings \cite{fisc}.

In what follows, we elucidate the distinct features of the ground state many-body properties of 
stably fragmented condensates by contrasting them with the fragility of symmetry-point-induced fragmented condensate states.
To this end, and to illustrate the salient features of stable fragmentation for interacting bosons in a most
transparent fashion, we use a simple model with just two macroscopically occupied field operator modes. Within this model, 
a closed analytical expression for fragmented condensate many-body states 
can be devised. 
Using the corresponding many-body amplitudes in Fock space, we demonstrate that a fragmented state with {\it continuous} (i.e., nonsingular) probability amplitudes for the Fock basis states is stable against quantum fluctuations of the occupation numbers of the modes and their relative phase, as well as against single-particle tunneling between the two states. We contrast this with the 
well-known instability of symmetry-point-induced fragmented states, which occurs in our model at vanishing pair-exchange coupling.
In addition, we investigate whether a (slightly) broken parity of orbitals significantly influences fragmentation.   
Finally, we discuss the perturbative effect of introducing an additional interacting mode. We find 
 that the single-particle density matrix has essentially 
 still two macroscopic eigenvalues, and the many-body state thus remains twofold fragmented.

   \section{two-mode fragmented states}
   \subsection{Hamiltonian and the many-body states}               
      We describe the quantum many-body phases of interacting bosons by the following two-mode Hamiltonian \cite{phil}, \begin{equation} \label{eq_1}
           \begin{split}
              \hat{H} & = \sum_{i = 0, 1} \left[ \epsilon_{i}\hat{n}_{i} + \frac{U_{i}}{2} \hat{n}_{i} \left( \hat{n}_{i} - 1 \right) \right] \\                                   
                      & + \frac{P}{2}(\hat{a}_{0}^{\dag} \hat{a}_{0}^{\dag}                                              \hat{a}_{1}\hat{a}_{1} +\mathrm{H.c.})                                                           + \frac{V}{2} \hat{n}_{0}\hat{n}_{1}.
           \end{split}
        \end{equation}
Without pair-exchange coupling, $P=0$, for $U_{0} + U_{1} - V > 0$, we obtain a Fock state $|N_{0}, N_{1}\rangle$, where the particle number in the ground-state mode $N_{0} = \frac{N}{2} - \frac{(U_{0} - U_{1})(N-1) + 2(\epsilon_{0} - \epsilon_{1})}{2(U_{0} + U_{1} - V)}$. Here, $N=N_{0} + N_{1}$ is the total number of particles. 
To obtain the generic features of the ground state for the Hamiltonian (\ref{eq_1}), we expand in a linear superposition of Fock states, $|\Psi \rangle = \sum_{l=0}^{N} \psi_{l}|l\rangle$ where $|l \rangle \equiv |N-l, l\rangle$ \cite{phil}. This Fock state expansion, by its definition, respects total particle number conservation, and the many-body correlations are encoded in the 
generally complex distribution vector $\psi_{l}=|\psi_l|\exp[i\theta_l]$, with amplitude $|\psi_l|$ and phase $\theta_l$. 
The distribution satisfies according to the Hamiltonian (\ref{eq_1}) the $N+1$ 
equations
      \begin{equation} \label{Ham_eq1}
        \langle l|\hat{H}|\Psi\rangle = E\psi_{l} = \frac{P}{2}(d_{l}\psi_{l+2} + d_{l-2}\psi_{l-2}) + c_{l}\psi_{l},
      \end{equation}
      where the coefficients $c_{l}=\epsilon_{0}(N-l) +\epsilon_{1}l + \frac{1}{2}U_{0}(N-l)(N-l-1) +\frac{1}{2}U_{1}l(l-1) +\frac{1}{2}V(N-l)l$ and $d_{l} =\sqrt{(l+2)(l+1) (N-l-1)(N-l)}$. 
Eqs.\,\eqref{Ham_eq1} decompose into two \textit{independent} sets of equations containing 
the even and odd $l$ sectors of $\psi_{l}$ only.
 The two corresponding ground states in the even and odd $l$ sectors are therefore \textit{degenerate} in the continuum limit 
 of $N\rightarrow\infty$. 
  
To represent the structure of the many-body wavefunction sufficiently far away from the singular symmetry point $P=0$, 
we employ the  spinor wavefunctions \cite{muel}
   \begin{equation} \label{spinor}
      |\theta, \phi\rangle = \frac{1}{\sqrt{N!}} (u \hat{a}_{1}^{\dag} + v\hat{a}_{2}^{\dag})^{N}                               |0\rangle,
   \end{equation}
   where the coefficients read $u = e^{-i\phi/2} \mathrm{cos}(\theta/2)$ and $v = e^{i\phi/2} \mathrm{sin}(\theta/2)$. 
Due to the even and odd $l$ sector degeneracy, the  weights of even and odd sector $\alpha$ and $\beta$, respectively, 
are arbitrary (subject to normalization of the wavefunction).
Upon investigating the structure of the binomially-expanded spinor wavefunction basis above,  taking into account 
the degeneracy of the even-odd $l$ sector, we can write an ansatz for the many-body wavefunction in the form    
      \begin{equation} \label{ansatz}
         \begin{split}
            |\Psi\rangle & = \alpha\sqrt{\frac{2}{N!}} \sum_{k=0}^{N/2} \mathrm{C}_{N}^{2k} (u \hat{a}_{0}^{\dag})^{2k} (v\hat{a}_{1}^{\dag})^{N - 2k} |0\rangle  \\
                         & + \beta\sqrt{\frac{2}{N!}} \sum_{k=1}^{N/2} \mathrm{C}_{N}^{2k-1} ( u \hat{a}_{0}^{\dag})^{2k-1}(v \hat{a}_{1}^{\dag})^{N - 2k+1} |0\rangle,
         \end{split}
      \end{equation}
      where the binomial coefficients are $\mathrm{C}_{N}^{k} = \frac{N!}{(N-k)!k!}$, $u = \exp[-i{\phi_{2}}/{2}] \mathrm{cos}(\theta/2)$, $v = \exp[i{\phi_{2}}/{2}] \mathrm{sin}(\theta/2)$, and $\phi_2$ here represents a phase difference
      between the $l$ and $l+2$ modes, defined such as to enter identically for even and odd $l$ sectors. 
   We assume for simplicity that $N$ is even.
      

The normalization of the wavefunction implies $|\alpha|^{2} + |\beta|^{2} = 1$.
The phase relation between even and odd Fock states is defined by writing $\alpha = |\alpha|\exp[-i\phi_{1}/{2}]$ and 
$\beta = |\beta| \exp[i{\phi_{1}}/{2}]$. The matrix elements of the single-particle density matrix in this state are 
$\langle\Psi | \hat{a}_{0}^{\dag} \hat{a}_{0}|\Psi \rangle = N_{0} = N \mathrm{cos}^{2}(\theta/2)$, $\langle\Psi |\hat{a}_{1}^{\dag} \hat{a}_{1} |\Psi\rangle = N_{1} = N \mathrm{sin}^{2}(\theta/2)$, and $\langle\Psi|\hat{a}_{0}^{\dag} \hat{a}_{1}| \Psi\rangle = N |\alpha||\beta| \mathrm{cos}(\theta/2) \mathrm{sin}(\theta/2) \mathrm{cos}(\phi_{1}) e^{i\phi_{2}}$. 
      
Using the wavefunction ansatz (\ref{ansatz}), the total energy per particle reads \cite{order1/N},  
      \begin{equation} \label{eq_2}
         \begin{split}
            \frac EN & = \frac{N}{2}\left[U_{0} + U_{1}  - 2P\mathrm{cos}(2\phi_{2}) -V\right] \mathrm{sin}^{4}(\theta/2) \\
                                             & + \left[\epsilon_1-\epsilon_0 
                                             + N\left(\frac{V}{2} - U_0 + P\mathrm{cos}(2\phi_{2})\right)  \right] \mathrm{sin}^{2}(\theta/2) \\
                                             & + \frac{U_0}{2}N + \epsilon_{0}. 
         \end{split}
      \end{equation}      
It is easily verified that, minimizing the above energy expression, we can recapture within one wavefunction ansatz \eqref{ansatz} the (continuum limit) observations made in \cite{phil} for the many-body ground states of the two-model \eqref{eq_1}. 
We will now these quantum phases for the parameter regime $U_0 + U_1 + 2|P| - V > 0$ in more detail. 
For a numerical verification of the wavefunction ansatz \eqref{ansatz}, see section \ref{numerics}.
      
      \subsection{Coherence properties}      
      The first-order coherence and degree of fragmentation $\mathcal F$ \cite{phil}, 
      corresponding to the ansatz (\ref{ansatz}) are, respectively, given by  
      \begin{equation} \label{fir_coh1}
         g_{1} = \frac{1}{2}\langle \hat{a}_{0}^{\dag}\hat{a}_{1} +                                                 \hat{a}_{1}^{\dag}\hat{a}_{0} \rangle = 2 |\alpha||\beta| \sqrt{N_{0}N_{1}} \mathrm{cos}                    (\phi_{1}) \mathrm{cos}(\phi_{2}),
      \end{equation}      
      and, using $\mathcal{F} = 1 - \frac{2}{N} \left| \frac{N}{2} - N_{1} \right|$, 
      \begin{equation} \label{deg_fra1}
         \begin{split}
            \mathcal{F} & = 1 - \frac{2}{N}\sqrt{|\langle \hat{a}_{0}^{\dag} \hat{a}_{1}                                       \rangle|^{2} + \left( \frac{N}{2} - \langle \hat{a}_{1}^{\dag}                                     \hat{a}_{1}\rangle \right)^{2}} \\
                        & = 1 - \frac{2}{N} \sqrt{4|\alpha|^{2}|\beta|^{2}N_{0}N_{1} \mathrm{cos}^{2}                                   (\phi_{1}) + \left( \frac{N}{2} - N_{1} \right)^{2}}.
         \end{split} 
      \end{equation}  
      For $P < 0$, the minimization of energy terms associated with the relative phase $\phi_{2}$ in Eq.(\ref{eq_2}), $P \mathrm{cos}(2\phi_{2})N_{0}N_{1}$, determines the phase to be either $\phi_{2}$ = 0 or $\pi$ (mod $2\pi$). Suppose that $\phi_{1} = 0$ (this is achieved for $P < 0$ provided that an infinitesimally small Josephson-type coupling between the levels is present, cf.\,\,the discussion in 
      Sec.\,\ref{Josephson} and Eq.\, \eqref{Jose1} below), and $\phi_{2} = 0$, the first-order coherence reads $g_{1} = \sqrt{N_{0}N_{1}}$, implying that the ground-state phase is a \textit{coherent} (single condensate) state. On the other hand, 
      when $\phi_2 = \pi$, the ground state is a \textit{$\pi$-phase coherent} ground state, for which the system favors negative first-order coherence. 
      
The second-order coherence function is defined by  $g_{2} = \frac{1}{2} \langle \hat{a}_{0}^{\dag}\hat{a}_{0}^{\dag}\hat{a}_{1} \hat{a}_{1} + \hat{a}_{1}^{\dag}\hat{a}_{1}^{\dag}\hat{a}_{0} \hat{a}_{0} \rangle$. Evaluating it by using the ansatz \eqref{ansatz} 
yields  
      \begin{equation} \label{sec_coh1}
                 g_2  = N^2 \mathrm{sin}^{2}(\theta/2)\mathrm{cos}^{2}(\theta/2) \mathrm{cos}(2\phi_{2}) \\
                  =  N_{0}N_{1}.
      \end{equation}
The second-order coherence $g_{2}$ is independent on the relative phase $\phi_1$ between the coefficients $\alpha$ and $\beta$ and macroscopic, 
i.e., ${\cal O}(N^2)$, implying that the ground-state is intrinsically \textit{pair-coherent}, 
by virtue of energy minimization. 
By contrast, the first-order coherence $g_{1}$ depends on $\alpha$ and $\beta$ and proper coherent states in our model exist for 
$\alpha = \beta = 1/\sqrt{2}$ only. Minimizing the total energy with respect to $\mathrm{sin}(\theta/2)$, the occupation number in the ``excited'' single-particle state reads 
      \begin{equation} 
       N_1=\langle \hat{a}_{1}^{\dag}\hat{a}_{1} \rangle = \frac{
       \epsilon_{0} - \epsilon_{1} 
       -\left( \frac{V}{2} - U_{0} -|P| \right)N}{U_{0}+ U_{1} +2|P| - V} , \label{N1}
      \end{equation} 
a formula also valid for $P > 0$.  
         
Turning to positive pair-exchange coupling, minimal energy requires $\phi_{2} = \pi/2$ or $-\pi/2$. This results in vanishing first-order and negative pair coherence, $g_{1} = 0$ and $g_{2} = - N_{0} N_{1}$. The ground state in the form of Eq.\,\eqref{ansatz} can be rewritten as follows: 
      \begin{equation}\label{gs2}
         |\Psi\rangle = \sum_{k = 0}^{N/2} e^{\pm i\frac{\pi}{4}(N - 4k)} \left[\alpha  f_{2k} |2k \rangle \pm i \beta f_{2k+1} |2k + 1 \rangle \right],
         \end{equation}
      where the real amplitude function $f_k$ is defined by 
      \begin{equation} f_{k} = \sqrt{\frac{2k! (N-k)!}{N!}} C_{N}^{k} [ \mathrm{cos}(\theta/2) ]^{k} [ \mathrm{sin} (\theta/2) ]^{N - k}, \end{equation}
      and the upper and lower sign stands for $\phi_2=\pi/2$ and $\phi_2=-\pi/2$, respectively.
      
      Macroscopic and negative pair coherence, vanishing first-order coherence, and, in particular, a finite degree of fragmentation 
      $\mathcal F$ characterize the many-body state for $P > 0$ in our model 
      and correspond to \textit{fragmented} ground states. 
  \begin{figure}[t]
	      \centering
		     \includegraphics[scale=0.52]{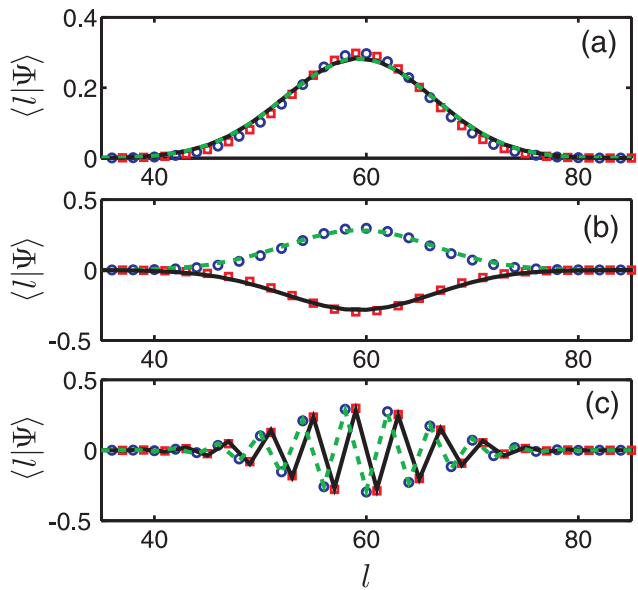} 
		     \caption{(color online) Comparison of numerics and the ansatz in Eq.\,\eqref{ansatz}. The coherent state (a) and $\pi$-phase coherent state (b) correspond to the parameters: $U_0 = V = 1$, $U_1 =0.8$, $P = -0.2$ while the fragmented state (c) is from $P = 0.2$ and identical other parameters to (a) and (b). $N=100$, $\Omega = 0$, and $\epsilon_{0} = \epsilon_{1} = 0$ for all cases here. Based on numerically solving Eq.(\ref{Ham_eq1}), the red squares indicate the even $l$ sector of $\psi_{l}$ and the blue circles the odd $l$ sector. The black solid line and the green dashed line correspond to the even and odd $l$ part of the analytic formula
		     for the distribution functions ${\mathcal P}(l)$, respectively.} 
	      \label{Fig1}
      \end{figure}   
         
      \subsection{Numerical verification of wavefunction ansatz}
       \label{numerics} 
      To verify the validity of our ground-state ansatz (\ref{ansatz}) for the two-mode model, we solve numerically the set of equations (\ref{Ham_eq1}) when $|\alpha| = |\beta| =1/\sqrt{2}$ to obtain $\psi_l$ and it compare it with 
the distributions ${\mathcal P}(l) \equiv \langle l |\Psi\rangle$,  
\bea
{\mathcal P}(l) &=& \sqrt{\frac{N!}{(N-l)! l!}}u^{N-l}v^{l}
\ea
for coherent and 
\bea
\begin{split}
{\mathcal P} (l) = \left\{ \begin{array}{cc} \sqrt{\frac{N!}{(N-l)! l!}} u^{N-l}v^{l} &\qquad \forall ~l~ \mathrm{even}\\ 
 \pm i\sqrt{\frac{N!}{(N-l)! l!}}u^{N-l}v^{l} &\qquad \forall ~l~ \mathrm{odd} 
\end{array}\right.
\end{split}
\ea 
for fragmented states, respectively.           
      
The results displayed in Fig.\ref{Fig1} show that the ansatz (\ref{ansatz}) is consistent with the numerical results based on Eq.\,\eqref{Ham_eq1}, establishing the validity of our generic ground-state expression in the interacting two-mode model. 

We now proceed to demonstrate the fundamental distinction between our fragmented states and  fragile fragmented states, by contrasting their respective responses to various perturbations, e.g.,  to quantum fluctuations and Josephson-type couplings between the single-particle states.

   \section{Stability of fragmented states against perturbations}
    
      \subsection{Quantum fluctuations of number and phase}

We start by defining states which infinitesimally differ from the ground state by writing the former in terms of the spinor states
$|\theta,\phi\rangle = |\theta_{0} + \delta, \phi_{2} +\phi \rangle$, where $\theta_0$ and $\phi_{2}$ determine respectively the particle number of the ground states in the two modes and the relative phase between $|l\rangle$ and $|l+2\rangle$. 

Expanding up to quadratic order in $\delta$ and $\phi$ around the ground state corresponding to 
the minimum of \eqref{eq_2}, and using the relation 
\eqref{N1}, we obtain that the low-lying excitations have the following 
energy, quadratic in phase ($\phi$) and number ($\delta$) fluctuations
         \begin{equation} \label{energy4}
            \begin{split}
               E(\theta_0 + \delta, \phi_{2} + \phi)- E(\theta_0, \phi_{2})  = 2|P|N_{0}
                                                           N_{1} \phi^{2} \\
                                                            +  \left[\left(U_{0} - 
                                                           \frac{V}{2} + |P| \right) \frac{3N_{0}
                                                           N_{1} - N_{0}^{2}}{4}\right. \\ 
                                                           \left. + \left(U_{1} - 
                                                         \frac{V}{2} + |P| \right) \frac{3N_{0}
                                                           N_{1} - N_{1}^{2}}{4}
                                                          \right. \\
                                                       \left.+ \left( \epsilon_{1} -  \epsilon_{0} 
                                                                   \right)
                                                           \frac{N_{0} - N_{1}}{4} \right]\delta^{2} .
            \end{split}
         \end{equation}
The crucial feature of the above result for the excited state energy is that it makes explicit that the energy of quantum fluctuations depends only on the {\it absolute value} of the pair-exchange coupling, $|P|$.
Therefore, the presently considered fragmented state with continuous distribution amplitudes $|\psi_l|$, obtained sufficiently far from  $P=0$ (also see below), is as robust against fluctuations of the $\psi_l$ distribution as a coherent state at the same value of $|P|$.  
Note in this respect that the factor $N_{0}N_{1}$ in front of the $\phi^2$ implies that the excitation energy per particle 
grows linearly in the total number of particles, so that the critical region of instability towards phase fluctuations 
around $P=0$ has the size $\delta P \sim {\cal O}(1/N)$ \cite{validity}.            
      
We now set the above discussion in relation to the analogous one for the well-known double well, as again described by a two-mode 
model \cite{milb}, with a Hamiltonian \bea
\hat H = - \frac{\Omega}{2}(\hat{a}_{L}^{\dag}\hat{a}_{R} + \mathrm{h.c.})
+\frac U2 \sum_{i=L,R}\hat n_i (\hat n_i -1) \label{Hdoublewell}
\ea 
in terms of the lowest-energy single-particle left/right eigenstates, putting 
$\epsilon_L=\epsilon_R=0$.  
The energy in terms of the spinor wavefunctions \eqref{spinor} reads
$E(\theta, \phi) = - \frac\Omega{2} N \cos\phi \sin\theta + U\left(\frac{N^{2}}{4}(                                      \mathrm{cos}^{2} \theta + 1) - \frac{N}{2} \right)$. 
The excitation energy around the two-mode coherent state, $|C \rangle = (\hat{a}_{0}^{\dag} + \hat{a}_{1}^{\dag})^{N}|0\rangle/\sqrt{2^{N}N!}$, in the double-well system therefore takes the form
   \begin{equation} \label{eq_6}
      E(\pi/2 + \delta, \phi) - E(\pi/2, 0) =  \frac{\Omega}{4} N\phi^{2} + \frac{N}{4}(\Omega + UN)\delta^{2} + \cdots,
   \end{equation}
   Comparing Eqs.(\ref{energy4}) and (\ref{eq_6}), a pronounced difference is manifest: The energy of phase fluctuations is associated with $N_{0}N_{1}\propto N^2$ for the fragmented many-body ground state of \eqref{eq_1}, while it is linear in the total particle number $N_{0} + N_{1}$ in the double-well system. 
   
Therefore, we come to the surprising conclusion that the single-trap fragmented state is {\it less} susceptible to phase fluctuations in the thermodynamic limit than its double-well counterpart,  provided single-particle- and pair-exchange amplitudes, $\Omega$ and $|P|$, for double-well and single trap respectively, are approximately of the same order. This is reflected in the width of the critical region around $P=0$ discussed in the above; there is no such critical region (critical in terms of the $N$-scaling of the exchange or tunneling coupling) for stability against phase fluctuations in the double-well case.

      \subsection{Josephson-type single-particle coupling}
\label{Josephson}
      A Josephson type perturbation of the form, 
             \begin{equation} \label{Jose1}
                \hat{H}_{J} = - \frac{\Omega}{2}(\hat{a}_{0}^{\dag}\hat{a}_{1} + \mathrm{h.c.}),
             \end{equation}           
             couples the two modes on the single-particle level. This can be due to tunneling in the case of a double well discussed      above. For a single trap, it can be realized by using two hyperfine states coupled by a two-photon Raman transition. 
             The energy of such a perturbation in terms of the state \eqref{ansatz} is 
      \begin{equation} \label{Jos_tun1}
         \begin{split}
            H_{J} & \equiv \langle \Psi|\hat{H}_{J} |\Psi\rangle \\
                  & = -\Omega N |\alpha||\beta|\mathrm{cos}(\theta/2)\mathrm{sin}(\theta/2) \mathrm{cos}(\phi_1) \mathrm{cos}(\phi_2),
         \end{split}
      \end{equation}   
Minimizing the Josephson energy $H_{J}$ yields $\alpha = \beta = 1/\sqrt{2}$ when $\Omega \mathrm{cos}(\phi_2) > 0$ and the two modes are both occupied macroscopically, i.e., $\mathrm{sin}(\theta/2) \neq 0$ and $1$. For a coherent state where $\phi_{2} = 0$, the Josephson-type energy is negative (the Josephson tunneling rate being positive definite, $\Omega>0$). 
Single-particle tunneling reduces the total energy and stabilizes the coherent state. By contrast, the $\pi$-phase coherent state ($\phi_{2} = \pi$) leads to the Josephson-tunneling energy $H_J$ being positive, which indicates that the $\pi$-phase coherent state is unstable
for any finite $\Omega$. 

When $\Omega = 0$, the coefficients $\alpha$ and $\beta$ in the even-odd superposition \eqref{ansatz}
can be any choice, subject to $|\alpha|^2+|\beta|^2=1$. However the Josephson-type tunneling, even when inifinitesimally small, pins down the explicit form of the pair-coherent states, in a similar manner to its establishing the conventional single-particle-coherent states. 
  
       \begin{figure}[tbp]
	      \centering
		     \includegraphics[scale=0.3]{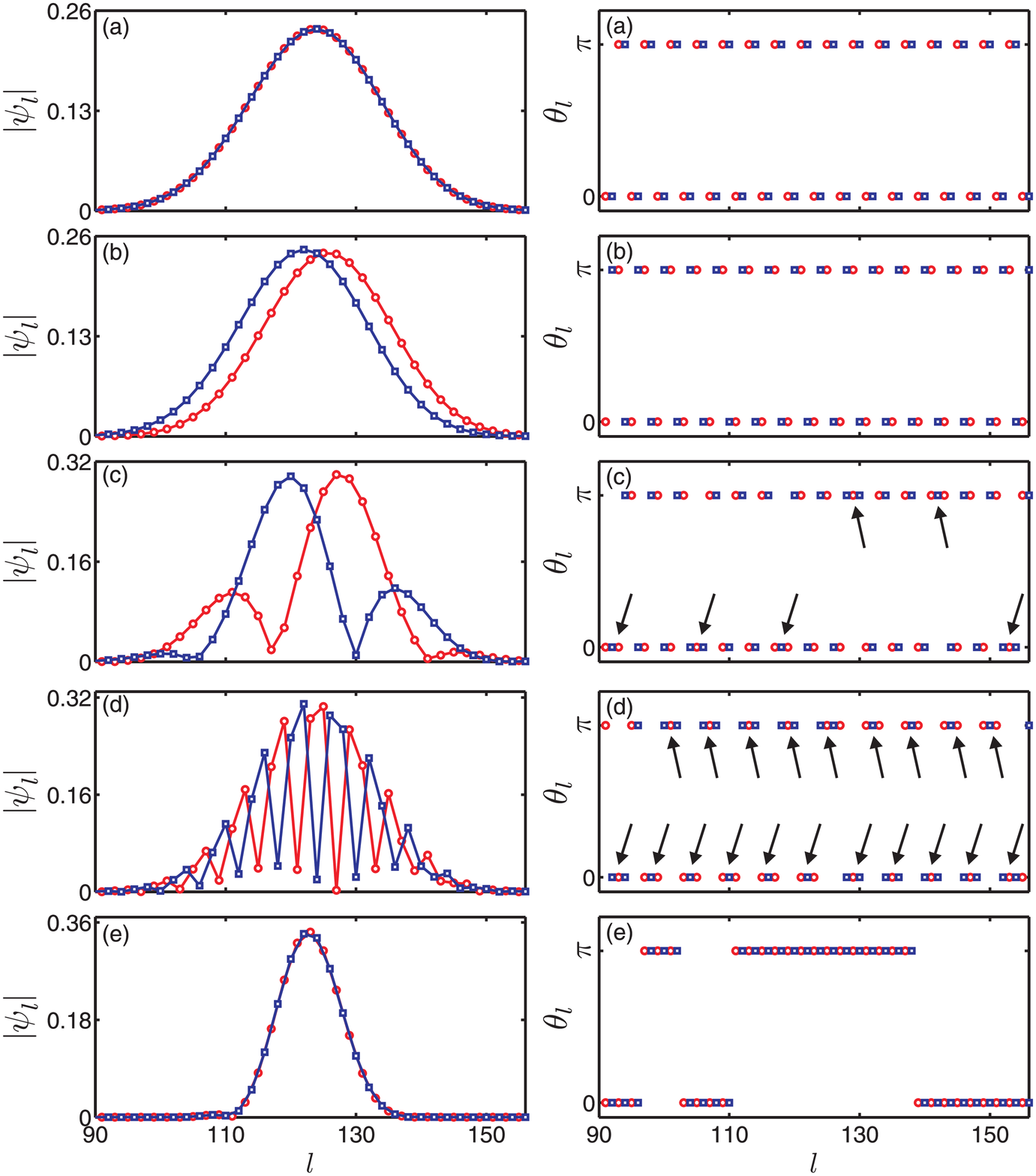} 
		     \caption{(color online) The Fock space distribution amplitude $|\psi_l|$ (left) and phase $\theta_l$		     (right) of the                                           ground-state wavefunction, varying the single-particle tunneling
		     $\Omega$, for a fragmented state with $P = 0.4$, $U_{0} = V = 1$, $U_{1} = 2/3$, $\epsilon_{0} = \epsilon_{1} = 0$ and $N =                                        200$.
		     The Josephson-type coupling increases from top to bottom, $\Omega =                                         0$ (a), $0.015NU_{0}$ (b), $0.1NU_{0}$ (c), $0.4NU_{0}$ (d), and                                   $0.8NU_{0}$ (e). 
		     Red circles represent the odd $l$ sector and blue squares the even $l$ sector. The arrows pointing to the phase data indicate the increasing breakup of the phase structure of the fragmented state. The phase structure of the ground state of \eqref{eq_1}, i.e., for $\Omega=0$, alternates according to the scheme $(0, 0, \pi, \pi, 0, 0, \pi, \pi, \ldots)$.} 
	      \label{Fig2}
      \end{figure}      

      \begin{figure}[btp]
	      \centering
		     \includegraphics[scale=0.3]{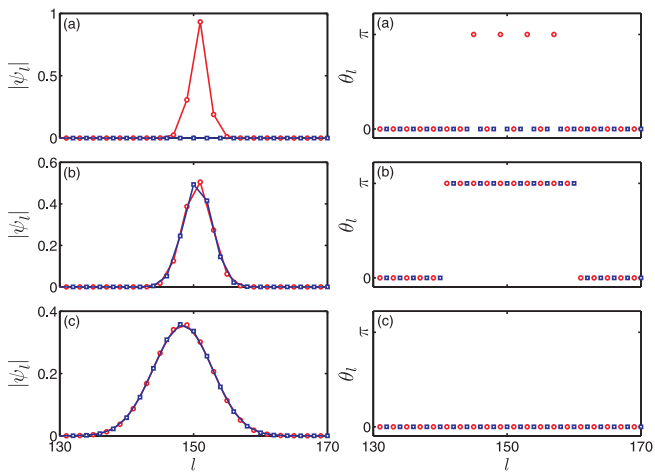} 
		     \caption{(color online) Evolution of a fragmented state which is ``almost'' a Fock state 
		     upon variation of the single-particle tunneling coupling $\Omega$.  
 We use 
parameters $P = 0.0001$, $U_{0}= V = 1$, and $U_{1} = 2/3$. In (a), $\Omega=0$, while 
 plots  (b) and (c) represent the response to $\Omega = 0.003NU_{0}$ and $0.015NU_{0}$.} 
	      \label{Fig3}
      \end{figure}

      As mentioned in the Introduction, the fragmented many-body states found in previous work 
emerge at special symmetry points of the Hamiltonian in question, 
and are either macroscopically occupied single Fock states \cite{muel}, or Schr\"{o}dinger-cat states consisting of the coheent superposition of macroscopically distinct single Fock states \cite{ho1}. For the latter, the instability of this type of (maximally) fragmented state is manifest as small quantum fluctuations rapidly destroy fragile superpositions of the NOON type.

To illustrate and compare the stability features of both robust and Fock-state-type fragmented states, we produce them 
within our two-mode model by adjusting the pair-exchange coupling $P$ close to zero, and then examine their stability against Josephson-type perturbations.

As seen in Fig.\,\ref{Fig2}\,(a) and (b), for a small $\Omega = \mathcal{O}(0.01NU_{0})$, the robust fragmented state with Gaussian-shaped distribution experiences only a small alteration: The distribution of even and odd $l$ parts of $|\psi_{l}|$ slightly shifts relative to each other while the phase structure ($\theta_l$), in particular the crucial feature of $\pi$ phase jumps between even and odd $l$, remains unchanged. 
With increasing single-particle tunneling, for example, from $0.1NU_{0}$ (c) to $0.8NU_{0}$ (e), the smooth amplitude function 
develops increasing modulations and the corresponding phase structure is broken gradually due to the competition between single-particle tunneling and pair-exchange coupling.
Finally, a uniform phase is established and the Gaussian distribution of the many-body wavefunction revives, when the ground-state properties are largely dominated by a very large single-particle tunneling of order the interaction energy ($\Omega = \mathcal{O}(NU_{0})$; see Fig.\ref{Fig2} (e)). 

By contrast, a comparatively small Josephson-type tunneling (for example, $\Omega = \mathcal{O}(0.01NU_{0})$) breaks a fragile fragmented state
in the critical region around $P=0$ (Fig.\ref{Fig3} (a)) and the system is driven towards a coherent state (Fig.\ref{Fig3} (b)); the sharply peaked distribution and non-uniform phase structure for this fragile fragmented state rapidly evolves into a Gaussian distribution and a uniform phase structure with increasing $\Omega$.

\begin{figure}[tbp]
	      \centering
		     \includegraphics[scale=0.39]{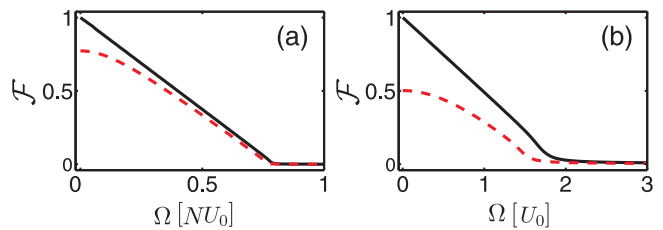} 
		     \caption{(color online) Variation of the degree of fragmentation with tunneling coupling 
		      for a robust ($P = 0.4$ (a)) and a fragile fragmented state with $P = 0.001$ ($=\mathcal{O}(1/N)$, (b)).  
		      Black: $U_0 = U_1 = V = 1$. Red: $U_0 = V =1$ and $U_1 = 2/3$; $\epsilon_0 = \epsilon_1 =0$ and $N = 1000$.
		      Note the macroscopically distinct scales on the $\Omega$ axis in (a) and (b).} 
		      \label{variation} 
	      \label{Fig4}
      \end{figure}

We summarize these properties in  
Fig.\,\ref{Fig4}, which shows the variation of the degree of fragmentation upon increasing the single-particle tunneling, and thus the conversion from a fragmented condensate to a single condensate.    
The Fig.\ref{Fig4}\,(b) demonstrates that the (close to) Fock-like fragmented states 
quickly decay into a coherent state already for small $\Omega$.  
On the other hand, the fragmented states with large distribution width ($P = 0.4$) are persistent, 
and a comparatively huge $\Omega$ ($\mathcal{O}(NU_{0})$), i.e., of order
the interaction energy scale, is necessary to transform them to a coherent, single condensate, state. 
 
\subsection{Modes without definite parity}
 In general, there are interaction-induced terms of the form  
 $-\frac12 J_2 \hat a_0^\dagger \hat a_0^\dagger \hat a_0 \hat a_1+{\rm h.c.}
 =-\frac12 J_2\hat n_0 \hat a_0^\dagger \hat a_1 +{\rm h.c.}$,  
 as well as $-\frac12 J'_2 \hat a_1^\dagger \hat a_1^\dagger \hat a_1 \hat a_0+{\rm h.c.}= -\frac12 J'_2\hat n_1 \hat a_1^\dagger \hat a_0 +{\rm h.c.}$ in addition to those occurring in \eqref{eq_1}. 
This happens even in the presently considered case of a single trap, when the modes do not have a definite parity which is different for the two modes.
On the other hand, when the modes respect a definite parity, the coefficients $J_2=\int\int dx dx' V(x-x')  (\psi^*_0(x))^2\psi_0(x')\psi_1(x')$ and $J'_2=\int\int dx dx' V_{\rm int}(x-x') (\psi^*_1(x))^2\psi_1^*(x')\psi_0(x')$ are zero, where
$V_{\rm int}(x-x')$ is the two-body interaction, assumed to be itself of a definite parity.
The terms $\propto J_2,J'_2$ lead to number-weighted tunneling-coupling processes.
The corresponding weight in the energy scales with $N^2J_2$ and $N^2J'_2$; hence, when $J_2$ and $J'_2$ respectively 
are of the same order as $P$, these terms will have a significant influence on the degree of fragmentation $\cal F$, similar to 
a tunneling rate $\Omega$ of order $NU_0$.   

\begin{figure}[t]
	                   \centering
		                 \includegraphics[scale = 0.375]{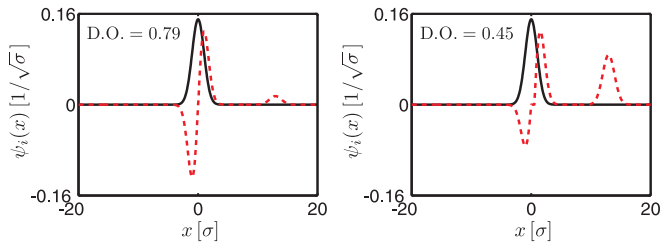} 		        	     
		                 \caption{(color online) Examples for the deformation of the first excited state of the harmonic oscillator $\psi_{1}(x)$ (red dashed line) away from exact odd parity with varying degree of overlap. The ground state $\psi_{0}(x)$ is shown by the black solid line.
		                 The excited state with definite (odd) parity here corresponds to D.O.$=0.8$.} 
	                    \label{Fig1Reply}
                   \end{figure}

We demonstrate now by a specific example that, in a single trap, we do not expect number-weighted tunneling to play a 
significant role when the parity of the modes is weakly broken. 
We take as the two (real) modes the ground state of the harmonic oscillator $\psi_0(x)=(\pi \sigma^{2})^{-1/4} \exp[-x^{2}/2 \sigma^{2}]$ and the first excited state $\psi_1(x)=\left({\sqrt{\pi} \sigma^{3}/2}\right)^{-1/2} x \exp[- x^{2}/2           \sigma^{2}]$ in one dimension, and deform the mode $\psi_1(x)$ away from odd parity, by introducing,  
at an arbitrary position where $\psi_0(x)$ has no weight, an additional maximum of $\psi_1(x)$ (cf.\,Fig.\,\ref{Fig1Reply}), 
keeping in the process exact orthogonality, $\int dx \psi_0(x) \psi_1(x)=0$, and normalization of the modes. 
To parametrize the change in the orbital's shape away from definite parity, we define the degree of orbital overlap as D.O.$\equiv\int dx |\psi_0(x) \psi_1(x)|$. When the degree of overlap tends to zero,
and the parity violation becomes large, we effectively have the familiar case of a double-well potential, see   
Fig.\,\ref{Fig1Reply}, which displays examples for the excited state orbital when 
deformed away from the definite parity state corresponding to the first excited state of the harmonic oscillator.  
 
\begin{figure}[b]
	                   \centering
		                 \includegraphics[scale = 0.45]{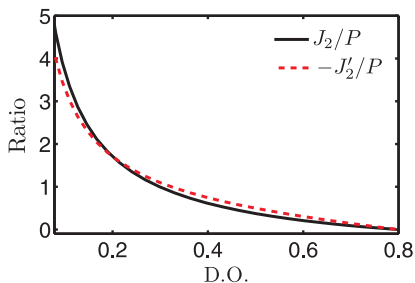} 		        	     
		                 \caption{(color online) Ratios of coupling constants $J_{2}/P,-J'_2/P$, as a function 
		                                         of the degree of overlap of the orbitals. The degree of parity breaking increases
		                                         from right to left. The ratios are identical, 
		                                         e.g., for a contact 
		                                         $V_{\rm int}=g\delta (x-x')$ or a dipolar interaction $V_{\rm int}=3g_{\mathrm{d}}/4\pi |x-x'|^{3}$.} 
	                    \label{Fig2Reply}
                   \end{figure}                   
We then compute the coefficient ratios $J_2/P, J'_2/P$ for varying degree of overlap. From Fig.\,\ref{Fig2Reply},   
we conclude that as long as the degree of overlap D.O. remains large and parity of the modes is almost preserved, 
the fragmentation remains largely unaffected, while for larger $|J_2/P|,|J'_2/P|$, the effect of interaction-induced 
number-weighted tunneling on fragmentation will be, as expected, equivalent to a tunneling rate of order $NU_0$, cf. Fig.\,\ref{variation}\,(a). 
As long as parity is only weakly broken, and $|J_2|\ll |P|$, stable fragmentation is thus still determined  
by the finite value of $P$. Weak breaking of parity we expect, for example, when the trap confining the bosons does itself  
not exactly respect a definite parity.
On the other hand, for maximally broken parity, pair-exchange becomes less important than number-weighted tunneling, 
and the physics of an effective double well takes over.
\subsection{Adding a third mode}   
      We now study whether two-mode fragmented states 
      continue to exist  
      when a third mode, potentially also macroscopically occupied, couples to the two modes. 
The three-mode Hamiltonian is in analogy to the two-mode case given by  
      \begin{equation} \label{3mode_eq1}
         \begin{split}
            \hat{H} & = \sum_{i=0,1,2}\left[\epsilon_{i}\hat{n}_{i} + \frac{U_{i}}{2} \hat{n}_{i} (\hat{n}_{i} - 1) \right] \\
                    & + \frac{P_{0}}{2} (\hat{a}_{0}^{\dag}\hat{a}_{0}^{\dag}\hat{a}_{1} \hat{a}_{1} + \mathrm{H.c.}) + \frac{P_{1}}{2} (\hat{a}_{1}^{\dag}\hat{a}_{1}^{\dag}\hat{a}_{2} \hat{a}_{2} + \mathrm{H.c.})   \\
                    & + \frac{P_{2}}{2} (\hat{a}_{0}^{\dag}\hat{a}_{0}^{\dag}\hat{a}_{2} \hat{a}_{2} + \mathrm{H.c.}) + \frac{V_{0}}{2}\hat{n}_{0}\hat{n}_{1} \\
                    & + \frac{V_{1}}{2}\hat{n}_{1}\hat{n}_{2} + \frac{V_{2}}{2} \hat{n}_{0} \hat{n}_{2},
         \end{split}
      \end{equation}   
      where additional pair-exchange terms associated with  the couplings $P_{1}$, $P_{2}$ as well as density-density
      type terms $\propto V_1,V_2$ are included.
      We set $\epsilon_{i} = 0$ for simplicity in what follows. 
      The first-order coherence measures for the three modes are specified by
      \begin{equation} \label{3mode_eq3}
         g_{ij}^{(1)} = \frac{1}{2} \langle \hat{a}_{i}^{\dag} \hat{a}_{j} + \hat{a}_{j}^{\dag}                            \hat{a}_{i} \rangle,
      \end{equation} 
      where $i$, $j=\{0,1,2\}$ labels the modes; in particular $g_{01}^{(1)}\equiv g_1$ as defined in Eq.\,\eqref{fir_coh1}. 
      
      \begin{figure}[btp]
	      \centering
		     \includegraphics[scale = 0.5]{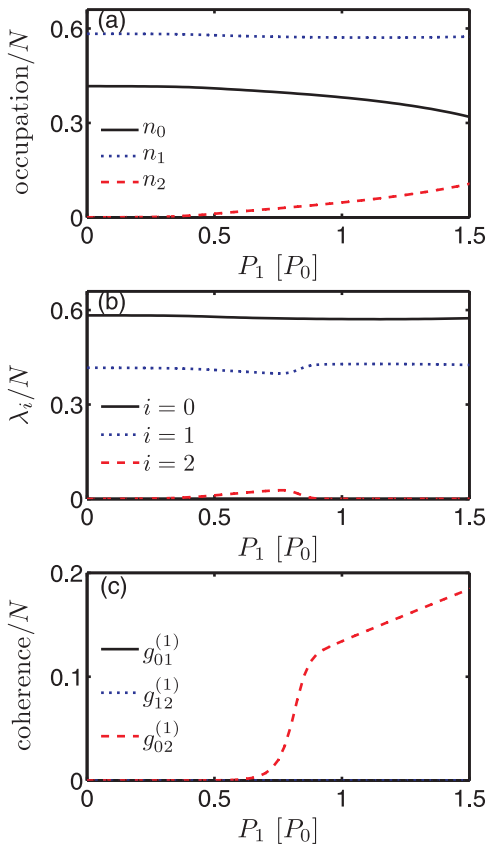} 
		     \caption{(color online) Illustration of the perturbative influence of an additional mode
		     on a twofold fragmented state. We display the average occupation of the three modes in (a),  the three eigenvalues $\lambda_i$ of the single-particle density matrix in (b), and the first-order coherence measures \eqref{3mode_eq3} between various modes (c), as a function of the pair-exchange tunneling $P_1$ 
		     between the modes $i=1$ and $i=2$ in units of $P_0$.   Here, $P_{1}$ $= P_{2}$ is fixed, as well as $U_{0} = 1$, $U_{1} = 0.8$, $U_{2} = 1.2$, $V_{0} = 1$, $V_{1} = V_{2} = 1.2$, 
		     and $N = 200$.} 
	      \label{Fig7}
      \end{figure}        
      We start from  a twofold fragmented state as a ground-state solution of the three-mode Hamiltonian above,  
with no (macroscopic) occupation in the third mode. 
To study the perturbative influence of the third mode, we then increase the pair-exchange couplings $P_{1}$ and $P_{2}$ simultaneously
from zero to the same order as $P_{0}$, such that the particle number in the third mode increases gradually cf.\,Fig.\,\ref{Fig7}\,(a).
We study the paradigmatic case shown in Fig.\,\ref{Fig7} 
because we anticipate, following the discussion in Sec. III B, that among the parameters associated with the third mode the pair-exchange parameters 
$P_{1}$ and $P_{2}$ influence the cohrence properties of the many-body state most significantly. 
With increasing $P_{1}= P_{2}$, both $\lambda_{2}$ and $n_{2}$ increase until a maximum of $\lambda_2$ occurs.
At the same time, this maximum indicates the transition to a novel two-mode fragmented state, for which a finite coherence between the modes $i=0$ and $i=2$ develops, see Fig.\,\ref{Fig7}\,(c). 
That is, moving away from the small maximum in $\lambda_2$ 
towards larger values of $P_{1}$, $\lambda_{2}$ decreases from its small peak to zero again, and the threefold fragmented state converts again into a fragmented state with just two macrosopic eigenvalues of the 
single-particle density matrix instead of three. 
We verified that this behavior qualitatively also occurs if we keep, e.g., $P_2=0$ and only increase $P_1$.   
Whether either macroscopic  $g_{02}^{(1)}$ or $g_{12}^{(1)}$ develops (limiting our discussion to $P_i>0$),  
 varies with the choice of parameters. 
 For example, 
choosing $U_0=0.8$ and $U_1=1$ in Fig.\,\ref{Fig7},
$g_{12}^{(1)}$ becomes macroscopic instead of $g_{02}^{(1)}$.

We thus conclude that the two-mode fragmented state can be stable against perturbations due to interaction coupling with a third 
mode, even when this mode also develops a significant (macroscopic) occupation. 

   \section{concluding remarks}
   We have investigated whether fragmented states in a two-mode model with pair-exchanges, representing 
   bosons interacting by two-body forces in a single trap, are robust against perturbations of various origin.
We constructed an analytical ansatz describing the many-body solutions, 
which was verified numerically. Concentrating on a fragmented condensate solution originating from positive pair-exchange coupling, we reveal its persistence against quantum fluctuations of particle numbers in the modes and their relative phase, 
against single-particle tunneling and (weakly)
broken parity of the modes,  
as well as the introduction of an additional interacting mode.   
A possible extension of the present work is to determine the spatial orbitals with multiconfigurational methods \cite{alon,grond,kangsoo}, and hence to solve the stability problem self-consistently. For example, fragmented condensates have been found in the crossover from a single condensate to ``fermionization'' \cite{Girardeau}, for tightly laterally confined bosons \cite{LL,Fischer}. It would, then, be of 
interest to investigate the more generic situation %
that also the coupling coefficients 
in the Hamiltonian can (in principle significantly) vary upon perturbing the system, 
and the ensuing consequences for the stability properties of fragmented condensates. 
 
One might legitimately ask whether rising temperatures above absolute zero, as $T=0$ was assumed in what precedes, destroy  the fragmented condensate states.
      In the canonical ensemble, the thermal average of the operators 
      occurring in the degree of fragmentation as defined in \eqref{deg_fra1}  
      is given by $\langle \hat{\mathcal{O}} \rangle = \sum_{\gamma=0}^{N} \frac{e^{- \frac{E_{\gamma}}{T}}}{Z} \langle E_{\gamma} | \hat{\mathcal{O}} |E_{\gamma} \rangle$, where the canonical partition sum $Z = \sum_{\gamma=0}^{N} e^{-E_{\gamma}/T}$ and 
$|E_{\gamma} \rangle$ are the eigenstates at energy $E_{\gamma}$.
We have verified that increasing the temperature to very large values $\ord(N U_0)$ does not change the degree of fragmentation $\mathcal F$ significantly; there is only a slight change (on the subpercent level) for the parameter values used in Fig.\,\ref{Fig2}. The fragmentation considered herein is therefore also not sensitive to finite temperature effects.

In summary, the present study demonstrates that there exist robust fragmented-condensate many-body states in a single trap, 
which share many features as regards their stability with single-condensate states, for which it 
is well established that they are stable under (sufficiently small) perturbations.
      
   \section*{ACKNOWLEDGMENTS}
      This research was supported by the NRF Korea, grant Nos. 2010-0013103 and 2011-0029541.


\begin{thebibliography}{000}
\bibitem{order1/N} Note that we neglect in \eqref{eq_2} and in the following equations $\ord(1/N)$ terms, however fully keep the single-particle energies, assuming their effective $N$-scaling to be in general adjustable.
\bibitem{validity} The ansatz \eqref{ansatz} for the wavefunction, with definite $\phi_2$, is valid outside this critical region, i.e.,    when $P\gtrsim \ord(1/N)$.  At $P=0$, an equal-weight average over $\phi_2$ yields an applicable wavefunction ansatz
   (corresponding to a single two-mode Fock state), while away
   from $P=0$, a (Gaussian-)weighted  average of \eqref{ansatz} 
   may be employed analogous to the one performed in \cite{muel} for the single-particle 
   tunneling Hamiltonian in Eq.\,\eqref{Hdoublewell}.
\end{thebibliography}
\end{document}